\documentclass[fleqn,usenatbib]{mnras}

\usepackage{newtxtext}

\usepackage[T1]{fontenc}

\DeclareRobustCommand{\VAN}[3]{#2}
\let\VANthebibliography\thebibliography
\def\thebibliography{\DeclareRobustCommand{\VAN}[3]{##3}\VANthebibliography}

\usepackage{graphicx}	
\usepackage{amsmath}	
\usepackage{amssymb}	
\usepackage{url}
\usepackage{ulem}

\title{J2102$+$6015: a young radio source at $z= 4.575$}

\author[Yingkang Zhang et al.]{
Yingkang Zhang$^{1}$\thanks{E-mail: ykzhang@shao.ac.cn},
Tao An$^{1}$,
S\'andor Frey$^{2,3}$\thanks{E-mail: frey.sandor@csfk.org},
Xiaolong Yang$^{1}$,
M\'at\'e Krezinger$^{4,2}$,
Oleg Titov$^{5}$,
\newauthor
Alexey Melnikov$^{6}$,
Pablo de Vicente$^{7}$,
Fengchun Shu$^{1}$,
and
Ailing Wang$^{1,8}$
\\
$^{1}$Shanghai Astronomical Observatory, Chinese Academy of Sciences, 80 Nandan Road, Shanghai 200030, China\\
$^{2}$Konkoly Observatory, ELKH Research Centre for Astronomy and Earth Sciences, Konkoly Thege Mikl\'os \'ut 15-17, H-1121 Budapest, Hungary\\
$^{3}$Institute of Physics, ELTE E\"otv\"os Lor\'and University, P\'azm\'any P\'eter s\'et\'any 1/A, H-1117 Budapest, Hungary\\
$^{4}$Department of Astronomy, ELTE E\"otv\"os Lor\'and University, P\'azm\'any P\'eter s\'et\'any 1/A, H-1117 Budapest, Hungary\\
$^{5}$Geoscience Australia, PO Box 378, Canberra, ACT 2601, Australia\\
$^{6}$Institute of Applied Astronomy, Russian Academy of Sciences, Kutuzova Embankment 10, St. Petersburg, 191187, Russia\\
$^{7}$Observatorio de Yebes (IGN), Apartado 148, 19180 Yebes, Spain\\
$^{8}$University of Chinese Academy of Sciences, 19A Yuquan Road, Shijingshan District, 100049 Beijing, China
}

\date{Accepted XXX. Received YYY; in original form ZZZ}

\pubyear{2021}

\begin{document}
\label{firstpage}
\pagerange{\pageref{firstpage}--\pageref{lastpage}}
\maketitle

\begin{abstract}
Jets of high-redshift active galactic nuclei (AGNs) can be used to directly probe the activity of the black holes in the early Universe. 
Radio sources with jets misaligned with respect to the line of sight are expected to dominate the high-redshift AGN population.
In this paper, we present the high-resolution imaging results of a $z = 4.57$ AGN J2102$+$6015 by analyzing its multi-epoch dual-frequency very long baseline interferometry (VLBI) data. The 8.4-GHz VLBI images reveal two major features along the east--west direction separated by $\sim10$ milli-arcsec (mas). From the spectral index map, both features show flat/inverted spectra. The separation between the two features remains almost unchanged over an observation period of $\sim13$~years, placing an upper limit of the separation speed as about 0.04~mas\,year$^{-1}$. Previous studies have classified the source as a GHz-peaked spectrum quasar. Our results indicate that J2102$+$6015 is most likely a young, compact symmetric object rather than a blazar-type core--jet source.
\end{abstract}



\section{Introduction}
\label{sec:intro}

Active galactic nuclei (AGNs) are the most luminous  persistent radiation sources in the Universe. They can be observed throughout almost the entire cosmic history from very high to low redshifts.
Radio-loud AGNs are an important sub-class of the population. Their radio emission is dominated by synchrotron radiation from the relativistic jets which are closely related to the accretion of their central supermassive black holes (SMBHs) \cite[e.g. ][]{1982MNRAS.199..883B,1984ARA&A..22..319B}. 
Radio-loud AGNs represent only $\sim~10$ per cent of the population. This percentage becomes smaller when moving to high ($z \ga 3$) redshifts \cite[e.g. ][]{2011MNRAS.416..216V}, while the reason for this deficit is still under debate.

At high redshifts, where most of the detected radio-loud AGNs show compact radio structure, selection effects should play an important role in the emergence of the radio sources. The larger the distance, the more luminous objects are likely to be observed because of the limited sensitivity of our instruments. Thanks to the technique of very long baseline interferometry (VLBI), the compact inner jet structure of high-redshift radio AGNs can be imaged with milli-arcsec (mas) resolutions \cite[e.g. ][]{1997A&A...325..511F,2008A&A...484L..39F,2010A&A...524A..83F,2020A&A...643L..12S, 2021AJ....161..207M}.
High-redshift AGNs exhibiting core--jet structures in VLBI images are usually identified as blazars, characterised by highly beamed jets pointing close to the line of sight. Their emission is significantly brightened by the Doppler boosting effect \cite[e.g. ][]{2010A&A...521A...6V,2013MNRAS.431.1314F,2016MNRAS.463.3260C,2017MNRAS.468...69Z}. However, blazars should constitute the minority of the radio AGN population at any redshift and therefore a larger number of non-blazar (unbeamed) radio sources would be expected at high redshifts as well \citep[e.g.][]{2016MNRAS.461L..21G}.

J2102$+$6015 is a powerful radio quasar at redshift $z=4.575$ \citep{2004ApJ...609..564S}, one of the most intriguing non-blazar AGNs at high redshifts. Previous VLBI imaging of this source showed a compact structure confined within 4~mas at 2.3~GHz. At 8.6~GHz, the source was slightly resolved into three components extending along east-west direction and an additional weak component was detected around 10 mas west of the peak emitting region \citep{2018A&A...618A..68F}.
The entire source shows an inverted radio spectrum peaking at $\sim$1~GHz \citep{2017MNRAS.467.2039C}, reminiscent of gigahertz-peaked spectrum (GPS) sources \citep{1998PASP..110..493O}. The brightest VLBI component has a moderate brightness temperature of $\sim 4 \times 10^{10}$~K \citep{2018A&A...618A..68F}, smaller than expected from a Doppler-boosted jet emission. All these observed characteristics suggest J2102$+$6015 is a non-blazar source.

In this paper, we present a detailed study of the mas-scale radio morphology, spectral index distribution, and component proper motion in this peculiar source, based on an extensive collection of archival and new dual-frequency VLBI data.
Section~\ref{sec:obs} provides information on the VLBI observations and the data reduction procedure. Section~\ref{sec:result} presents the results. Section~\ref{sec:Discussion} discusses the nature of  the radio emission in J2102$+$6015 and Section~\ref{sec:sum} gives a summary of our findings.
\section{Observations and data reduction}
\label{sec:obs}
In this section, we present the VLBI observational data used in this paper and describe the calibration, imaging and model-fitting procedures. There are data from a total of 12 epochs that had been collected and analysed. The basic information on these data sets is presented in Table~\ref{tab:obs}.

\subsection{Sensitive VLBA imaging session}
\label{sec:vlba}

We observed a sample of high-redshift radio-loud quasars including J2102$+$6015 with the U.S. Very Long Baseline Array (VLBA; project code: BZ064), aiming at obtaining high-sensitivity high-resolution images of these sources, to better constrain their jet properties in combination with archival VLBI data. A detailed description of this sample and the observational results will be presented elsewhere (Zhang et al., in prep.). 

This VLBA session was carried out at the frequency of 8.4~GHz on 2017 February 5. The data were recorded on the VLBA Digital Downconverter (DDC) system with four 128-MHz basebands each having 256 spectral channels.  The left and right circular polarizations share the total bandwidth equally, using 2-bit sampling to provide a total sampling rate of 2048~Mbit\,s$^{-1}$. To optimise the $(u,v)$ coverage, each source was observed with several scans over a broad time range. J2102$+$6015 was observed in six scans, each lasting for 5~minutes, giving a total on-source time of 30~minutes. 
The BZ064 VLBA experiment data were correlated at the DiFX correlator in Socorro, USA \citep{2011PASP..123..275D}, with 2~s integration time. Then it was downloaded to the China SKA Regional Centre \citep{2019NatAs...3.1030A} for further calibration and imaging. We used the NRAO Astronomical Image Processing System ({\sc AIPS}) software \citep{2003ASSL..285..109G} to calibrate the amplitudes and phases of the visibility data, following the standard procedure from the {\sc AIPS} Cookbook\footnote{\url{http://www.aips.nrao.edu/cook.html}}. 

The calibrated data were imported into the Caltech {\sc Difmap} package \citep{1997ASPC..125...77S} for imaging and model-fitting. Hybrid mapping procedures were performed in {\sc Difmap} \citep[e.g.][]{1980Natur.285..137R,1986AJ.....92..213L}, with several cycles of {\sc clean}ing and self-calibration. We present the final images by applying natural weighting to the visibility data. After mapping, we used the {\sc Difmap} program {\sc modelfit} to fit the self-calibrated visibility data with a series of circular Gaussian brightness distribution models, to identify and parameterise the VLBI components.

\subsection{Archival VLBI data sets}
\label{sec:astrogeo}

To study the time evolution of radio emission properties of J2102$+$6015 in detail, we also obtained archival data from the Astrogeo VLBI database\footnote{\url{http://astrogeo.org/}}, which is the largest collection of VLBI imaging data mostly based on a number of geodetic or astrometric projects \citep{2009JGeod..83..859P}, the VLBA Calibrator Surveys \cite[VCS; e.g.][]{2002ApJS..141...13B,2016AJ....151..154G}, and large astrophysical surveys of bright AGNs \cite[e.g.,][]{2005AJ....130.1389L,2020ApJS..247...57C}. 
The astrometric VLBI data from the Astrogeo archive are observed in a dual-band mode ($S/X$ bands, usually centred at 2.3 and 8.4~GHz frequencies) with typical integration time of a few minutes split into a few scans across long periods (from hours to days), leading to relatively low imaging sensitivity and adequate $(u,v)$ coverage. The visibility data obtained from the Astrogeo database have already been calibrated. Their imaging and model-fitting were done following similar steps as above in the {\sc Difmap} software (Section~\ref{sec:vlba}).

\subsection{Astrometric VLBI sessions}
\label{sec:rua}

We also collected 8.6-GHz VLBI data obtained during astrometric observations with an ad-hoc global VLBI array involving the Russian National Quasar VLBI Network \citep{SHUYGINA2019150}, the 40-m Yebes (Ys, Spain), and the 25-m Sheshan (Sh, China) radio telescopes, yielding a maximum baseline exceeding 9000~km. Seven epochs from 2017 to 2019 were used (under the project code series RUA, see Table~\ref{tab:obs}), occasionally combining close-epoch data sets for improving the imaging sensitivity, as was done by \citet{2018A&A...618A..68F}. 
Data from the RUA experiments were correlated at the DiFX correlator of the Institute of Applied Astronomy of the Russian Academy of Sciences in Saint-Petersburg, with 0.5~s integration time.
The $S/X$-band RUA experiments (Table~\ref{tab:obs}) used several $\sim16$-min scans distributed over 24-h periods, to improve the $(u,v)$ coverage with the relatively sparse VLBA array.
The data calibration, imaging and model-fitting were performed in {\sc AIPS} and {\sc Difmap}, through similar processing steps described in Section~\ref{sec:vlba}. More details of the observing setup and the data analysis can be found in \citet{2018A&A...618A..68F}.

\subsection{Brightness distribution modeling}
\label{sec:modeling}

From the series of {\sc clean} images, we found that the general appearance of the source did not change substantially during the period spanned by the observations, from 1994 to 2019. Therefore, we first generated a set of model components based on the best-quality data (experiment BZ064, Section~\ref{sec:vlba}), and then used them as initial parameters in the model fitting for other data sets in {\sc Difmap}. The interpretation of the multi-epoch images and brightness distribution models is presented in Section~\ref{sec:result}.

\section{Results}
\label{sec:result}

\subsection{Morphology}
\label{sec:morphology}

\begin{figure}
	\includegraphics[width=\columnwidth]{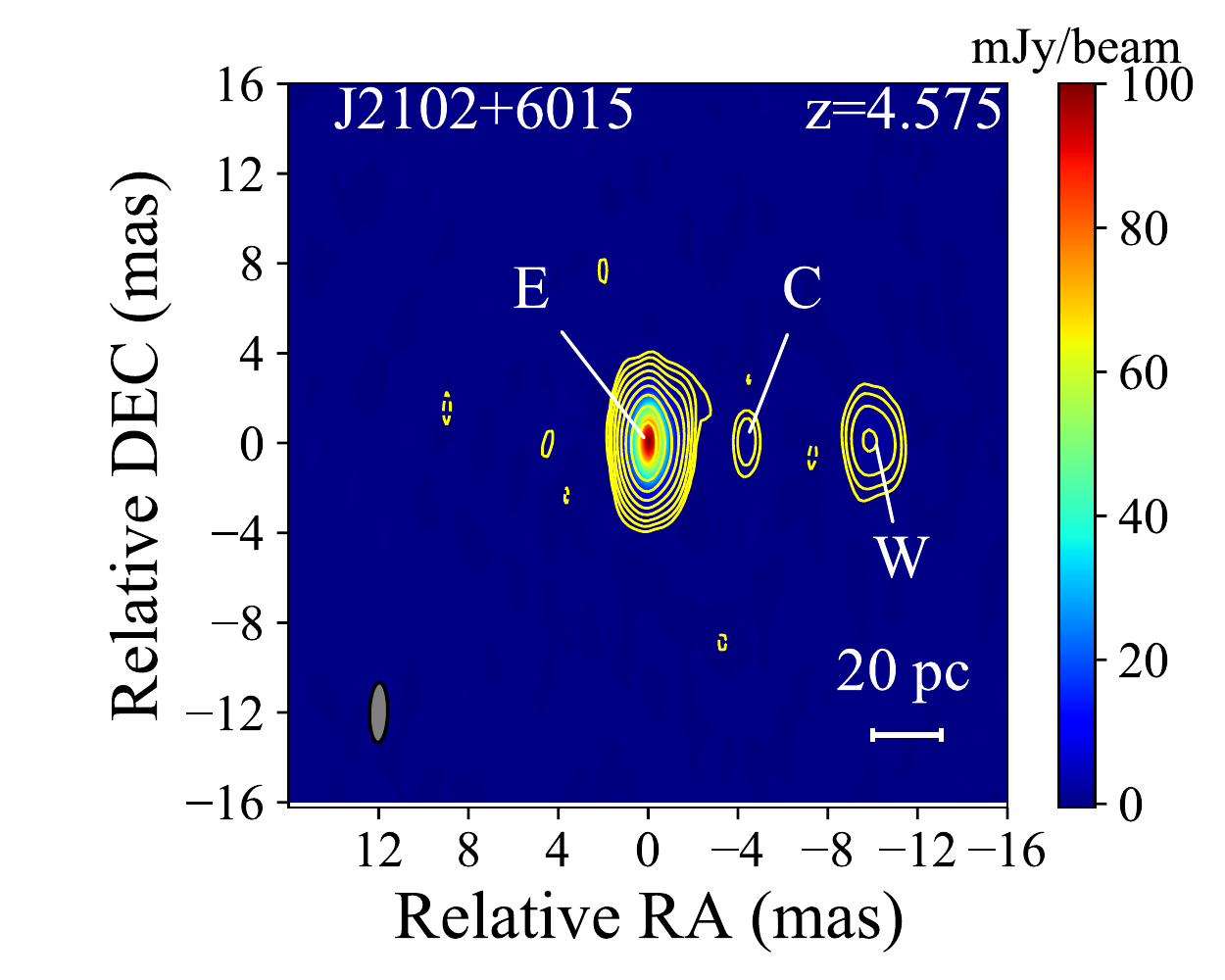}
    \caption{Naturally weighted high-resolution image of J2102$+$6015 created from the 8.4-GHz VLBA data. The lowest contours are $\pm 4$ times of the r.m.s. noise and a factor of 2 is used for contour increments.}
    \label{img:2017col}
\end{figure}

\begin{figure}
    \centering
	\includegraphics[height=20cm]{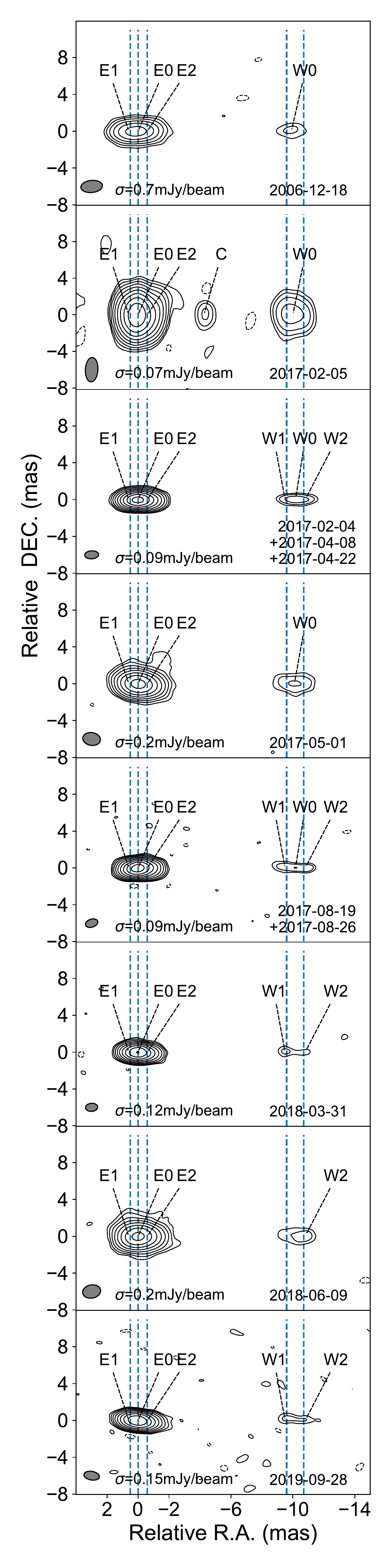}
    \caption{The demonstration of the component separation across all the available observing epochs for J2102$+$6015. The vertical dashed lines represent the positions of components E1, E0, E2, W1 and W2, from east to west. In each sub-image, the elliptical Gaussian restoring beam (FWHM) is shown in the lower-left corner and the lowest contours are at $\pm 3$ times of the r.m.s. noise ($\sigma$). The positive contour levels increase by a factor of 2. Note that to demonstrate the source components clearly, the images are stretched along right ascension (horizontal axis) with a factor of $~2$.}
    \label{img:demo}
\end{figure}

Figure~\ref{img:2017col} shows the image of J2102$+$6015 obtained from the 8.4-GHz VLBA observation in 2017. Three main features (labelled as E, W, and C) can be detected in the high-sensitivity {\sc clean} image. The bulk of the emission is concentrated in the component E which is the brightest at all epochs (Figure~\ref{img:demo}) with a peak intensity of $\sim 100$~mJy\,beam$^{-1}$. This feature is difficult to be fitted with a single circular or elliptical Gaussian model. Instead, E can be described well using three circular Gaussian models that are distributed along the east--west direction. We label these as E1, E0, and E2, from east to west.

A weak feature W is located at $\sim 10$~mas west of E0, corresponding to a projected distance\footnote{In a standard flat $\Lambda$CDM cosmological model with $H_{0}=70$~km\,s$^{-1}$\,Mpc$^{-1}$, $\Omega_{\mathrm m}=0.3$, and $\Omega_{\Lambda}=0.7$, 
at $z=4.575$, 1~mas angular size is converted to 6.55~pc projected linear size.
} of 65~pc. Among all available epochs, W is most prominent in the 2017 February 5 image which has a factor of 1.3--10 lower noise level than the other images (see Appendix~\ref{appendix}).
It was not detected in 1994 due to the relatively low sensitivity of the early VLBI image. At most epochs, the W feature appears resolved and can be modelled with two or three circular Gaussian components labelled as W1, W0, and W2 (Figure~\ref{img:demo}).
Feature C locates between the two main features E and W, and is only detected in the VLBA epoch on 2017 February 5, with a signal-to-noise ratio of $\sim$ 10.

\subsection{Radio spectrum}
\label{sec:radiospectrum}

\begin{figure}
    \centering
	\includegraphics[width=\columnwidth]{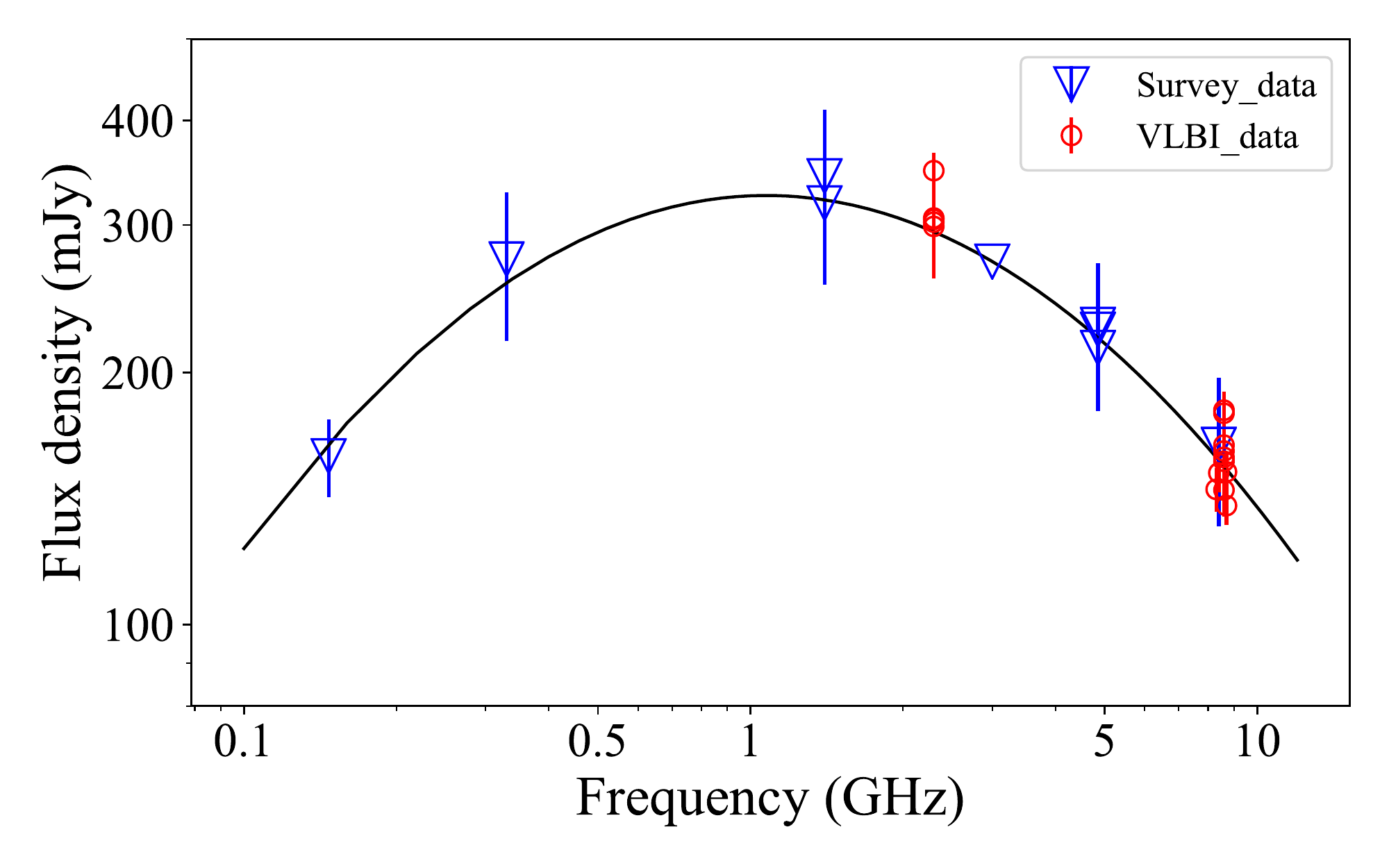}
    \caption{The radio spectrum of J2102$+$6015 constructed from VLBI data (red circles) presented in this paper and historical survey data (blue triangles). Compared to an earlier version \citep[Figure~25 in][]{2017MNRAS.467.2039C}, our spectrum is better constrained by the VLBI flux densities at 8~GHz. The spectrum is fitted with a log-parabolic function $\log(S_\nu)=a \times [\log(\nu)-\log(b)]^2 +c$, where $S_\nu$ is the flux density at the observing frequency $\nu$, and $a$, $b$, and $c$ are numerical parameters without direct physical meaning. The fit shows a peak at frequency $\nu_{0} = (1.07 \pm 0.08)$~GHz. The original data and references can be found in Table~\ref{tab:spec}.}
    \label{img:spec}
\end{figure}

J2102$+$6015 has been detected in a number of radio surveys from 150~MHz to 8.4~GHz and its radio spectrum was constructed by \citet{2017MNRAS.467.2039C}, showing a peaked spectrum. 
In addition to the data used by \citet{2017MNRAS.467.2039C}, we also added 2- and 8-GHz VLBI flux densities from the present study, and the 3-GHz data point from the Karl G. Jansky Very Large Array (VLA) Sky Survey (VLASS) \citep{2020PASP..132c5001L,2020RNAAS...4..175G}. 
We plot the flux density of J2102$+$6015 versus observing frequency in Figure~\ref{img:spec} and the corresponding data are listed in Table~\ref{tab:spec}. 
From the figures and tables, we found that the source does not show strong variability (i.e. $\left\lvert \frac{S_i-\mu}{S_i} \right\rvert \le 15$ per cent, where $S_i$ is the integrated flux density from each epoch and $\mu$ the average flux density calculated from all epochs in the same observing band) on timescales of years. 
Moreover, a comparison of the low-resolution VLA and high-resolution VLBI data taken at the same frequencies shows only small differences between the two, indicating that most of the total flux density originates from the compact $\sim 10$-mas scale radio structure imaged with VLBI (Figure~\ref{img:2017col}). 

Following \citet{2017MNRAS.467.2039C}, we used a log-parabolic function to fit the spectrum.
The fitted peak frequency is $\nu_{0} = (1.07 \pm 0.08)$~GHz, corresponding to $\sim 6$~GHz in the source rest frame. Our result agrees with that of \citet{2017MNRAS.467.2039C} within the uncertainties, because our multi-epoch VLBI measurements show stable flux density values during the 24-year time span and they are in good agreement with the low-resolution data. Therefore adding VLBI measurements does not significantly change the spectral fit. 
\subsection{Spectral index map}
\label{sec:spix}
Thanks to the simultaneous dual-band observations available from the Astrogeo database, a spectral index image can be produced to help reveal the source nature. We compiled the spectral index image based on the best-quality dual-band data on 2017 May 1.

Firstly, both images were produced with the same map sizes and cell sizes in {\sc Difmap}. The $X$-band image was then restored using the same beam with that of the lower-resolution $S$-band data. After that, the two images were aligned and the two-point spectral indices were calculated based on the intensity values pixel by pixel. According to {\sc modelfit}, the distance between E0 and W0 was about 1~mas larger at $X$ band than at $S$ band. This exceeds the positional accuracy of $\sim 0.5$~mas, and may be caused by either a frequency-dependent core shift (if it is core--jet source) or a synchrotron opacity effect in CSO hotspots \citep[e.g. ][]{2011A&A...535A..24S}.

\begin{figure}
	\includegraphics[width=\columnwidth]{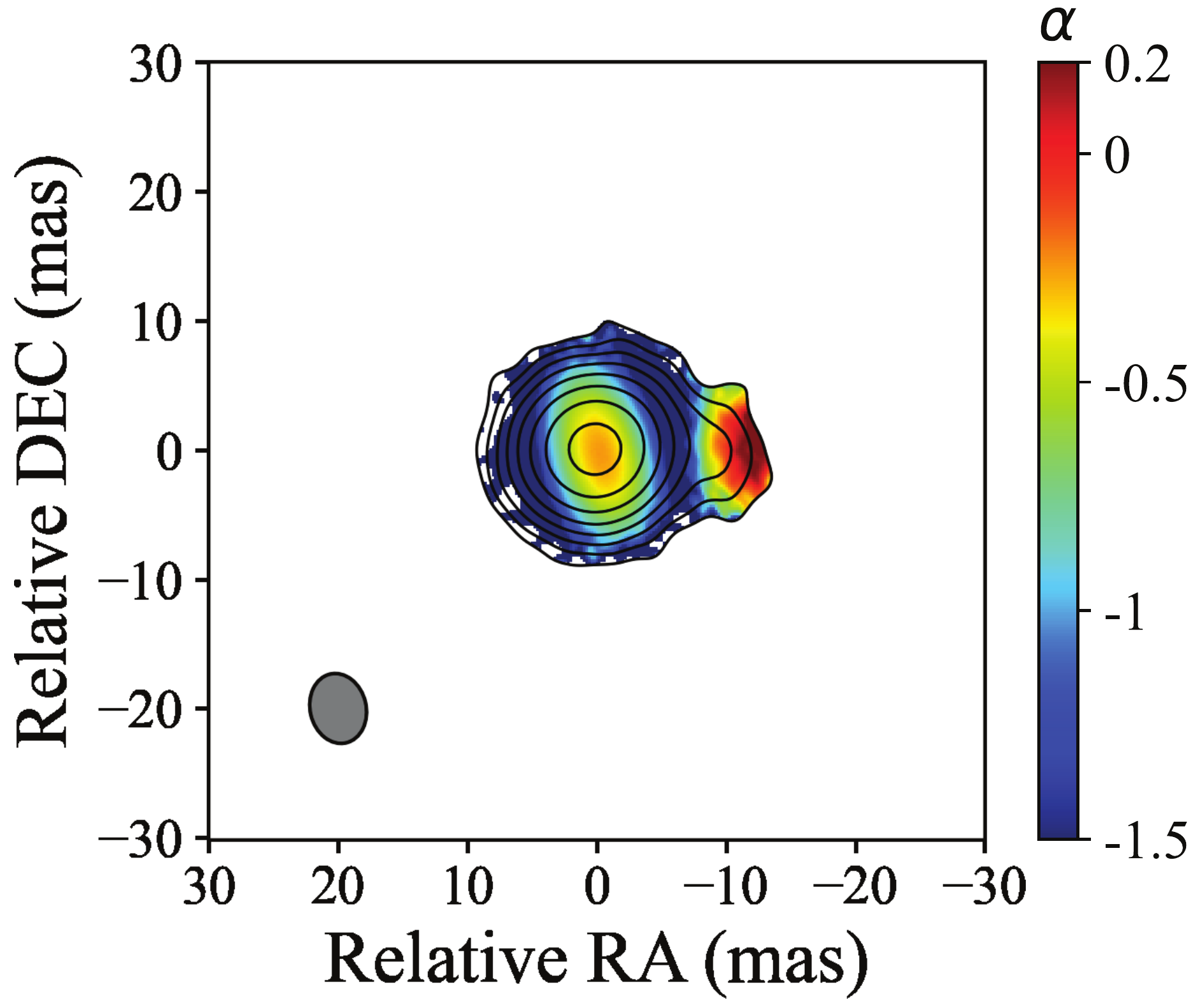}
    \caption{The two-point spectral index map of J2102$+$6015 derived from the 2.3- and 8.7-GHz VLBA images of 2017 May 1. The contours represent the 2.3-GHz map and the colours show the spectral index distribution.}
    \label{img:spix}
\end{figure}

Figure~\ref{img:spix} shows the spectral index map, where two components with flat or inverted spectra can be found around the positions E0 and W0. The image was made by aligning the positions of the E0 component at both bands. To avoid a bias from a different source nature, we also aligned the images at $S$ and $X$ bands based on W0 position, which led to a similar result with the eastern flat-spectrum component being close to E2.

For a synchrotron-emitting source, the flux density $S$ is proportional to $\nu^{\alpha}$, where $\nu$ is the observing frequency and $\alpha$ is the spectral index. In our spectral index map of J2102$+$6015 (Figure~\ref{img:spix}), two features are seen at the positions of E and W, with flat ( $\alpha_\mathrm{E} \approx -0.3$) and inverted ($\alpha_\mathrm{W} \approx 0.2$) spectra, respectively. 

\subsection{Proper motion, brightness temperature and luminosity}
\label{sec:propermotion}

The proper motion estimate of major components could provide a valuable addition to constraining the source properties. 
Figure~\ref{img:demo} exhibits 8-GHz images with all the fitted components at several epochs from 2006 to 2019. As mentioned in Section~\ref{sec:morphology}, E is the major feature in the source and can be fitted with three Gaussian components. The brightest component E0 is in the middle, with two adjacent components symmetrically located on its eastern (E1) and western (E2) sides. The much weaker western feature W is seen $\sim 10$~mas away from E0 and it seems to be resolved along the east--west direction in some images. The components (W1, W0 and W2) are marginally detected and resolved in some epochs, so that some of them could be below the detection threshold or involved with one another in the Gaussian model fitting.
Only in the VLBA image of 2017 February 5, component C is detected between E and W. Even though C could possibly be associated with a core in a Compact Symmetric Object \citep[CSO; ][]{1994ApJ...432L..87W}, it is too weak and detected only once. So we could not use C as a reference point for the proper motion measurements.

\begin{figure}
    \centering
	\includegraphics[width=0.8\columnwidth]{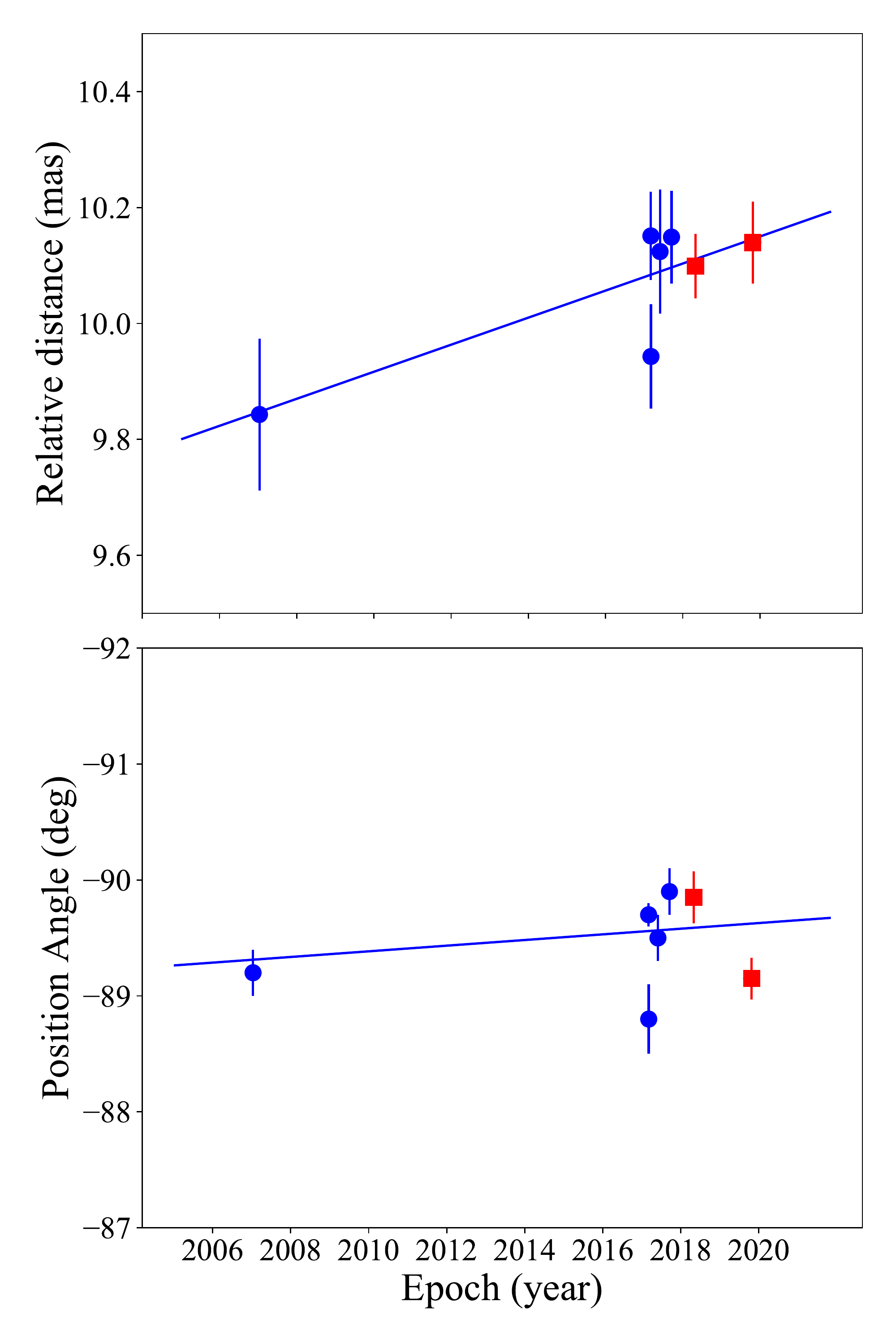}
    \caption{Change of the angular distance of W0 with respect to E0 (top), and the variation of the position angle (PA) of the W0 component as a function of time (bottom). The blue points are from the measurements of W0, the red squares represent the estimated values based on the positions of W1 and W2. The fitted proper motion is $(0.023 \pm 0.011)$~mas\,year$^{-1}$ and the PA change is $(-0.02 \pm 0.02)\degr$\,year$^{-1}$.}
    \label{img:pm}
\end{figure}

The E0 component is the most compact and the brightest in the source, and is detected in all epochs. Therefore we used E0 as a reference point to measure the change of W--E separation with time.
Component W0 could be used as the reference position for feature W, but it is detected only before 2018. We note that for the combined epochs from 2017 February--April and 2017 August, W0 is well within 1-$\sigma$ positional error of the centroid of W1--W2. We thus used the centroid positions based on W1 and W2 as reference positions for W in epochs 2018 March 31 and 2019 September 28. Therefore data from 7 epochs could be used for the apparent proper motion estimate, yielding a 13-year time span from 2006 to 2019.
We derived a linear proper motion of $(0.023 \pm 0.011)$~mas\,year$^{-1}$ (see Figure~\ref{img:pm}, top panel) for the feature separation. The position angle of W0 with respect to E0 remains practically constant during the period with minor variation within $1\degr$ (Figure~\ref{img:pm}, bottom panel). 
We estimated the brightness temperature of the dominant VLBI component (E0) using
\begin{equation}
T_{\mathrm B}=1.22 \times 10^{12} (1+z) \frac{S_{\nu}}{\theta_\mathrm{comp}^{2}\nu^{2}} \, {\mathrm K},
\end{equation}
where S$_{\nu}$ is the flux density of the VLBI component expressed in Jy, measured at $\nu$ GHz. $\theta_\mathrm{comp}$ is the full width at half-maximum (FWHM) diameter of the circular Gaussian component, in the unit of mas.
The estimated values are presented in the last column of Table~\ref{tab:parm}. In 1994, the outlier $T_\mathrm{B}$ value could be an artefact caused by the relatively low resolution. Since the E feature was not resolved into distinct components, the measured flux density was in fact the sum of E0, E1 and E2. By excluding this data point in 1994, we obtain an average brightness temperature of $T_\mathrm{B,E0} = (4.3 \pm 0.2) \times 10^{10}$\,K.

The monochromatic radio luminosity can be calculated as
\begin{equation}
\mathrm{L}_\nu = 4 \pi D_\mathrm{L}^2 \frac{S_\nu}{(1+z)^{1+\alpha}},
\end{equation}
where $D_\mathrm{L}$ is the luminosity distance and $\alpha$ is the spectral index. 
If we take $\alpha^{8.4}_{2.3} \approx -0.5$ from the partial fit to the spectral points in Figure~\ref{img:spec} and average the 8-GHz flux densities measured from all the $X$-band epochs, the monochromatic luminosity of the target source is $(13.8 \pm 1.3) \times 10^{27}$~W\,Hz$^{-1}$.

\section{Discussion}
\label{sec:Discussion}
High-resolution VLBI observations provide crucial pieces of information for the classification of J2102$+$6015. First of all, the high brightness temperature of the VLBI components ($>3 \times 10^8$~K) and the high radio luminosity ($\sim 10^{28}$ W Hz$^{-1}$) \citep[][this paper]{2016MNRAS.463.3260C,2018A&A...618A..68F} support a non-thermal synchrotron origin of the radio emission. The morphology, the surface brightness, and the spectral index of the E and W features are significantly different, ruling out the possibility of gravitationally lensed images. Here we report on the detection of component C at 8.4~GHz for the first time, and estimate the apparent proper motion of W with respect to E. In the following, we discuss these observational findings in the context of different conceivable scenarios for the nature of the radio source in J2102$+$6015.

\subsection{A quasar with core--jet morphology}
\label{sec:corejet}

The relatively high brightness temperature and compactness of E0 (Table~\ref{tab:parm}) would naturally lead us to regard it as the synchrotron self-absorbed AGN core. From the 8-GHz ($X$-band) images alone, J2102$+$6015 seems to resemble a usual core--jet source of a blazar in which E (actually E0) is the core and other weaker components C and W belong to a jet oriented towards the west. However, if E0 is the AGN core, then E1 can only be interpreted as a counter-jet component. 
This would be hard to reconcile with a general picture of apparently single-sided quasar jets caused by Doppler favoritism of the approaching (relativistically boosted) part.
Also, the spectral flattening of the W feature is difficult to interpret in the context of the core--jet structure, where an optically thin jet component is expected to have a steep power-law spectrum ($\alpha \le -0.5$). Another argument against E0 being the blazar core is that it does not show prominent flux density variability on year-long timescales. Finally, the brightness temperatures calculated for E0 (Table~\ref{tab:parm}) indicate Doppler boosting factors typically close to or below unity. Since the spectral peak is around 1~GHz (Figure~\ref{img:spec}), these brightness temperatures could be somewhat underestimated. However, the values measured for J2102$+$6015 at $\sim 2$~GHz are similar \citep[$\sim$ 4 $\times$ 10$^{10}$ K in][]{2018A&A...618A..68F}, therefore there is no evidence for strong relativistic beaming.

Another possibility is that the westernmost component W2 is the optically thick core. This would be consistent with its inverted spectrum. In this case, E would be a feature along the jet that is Doppler-enhanced due to bending towards the observer caused by the interaction between the jet and the interstellar medium (ISM) \citep[e.g.][]{2020NatCo..11..143A}.

\subsection{Binary supermassive black hole}
\label{sec:smbhb}
An intriguing possibility is that we see a supermassive black hole binary (SMBHB) system in J2102$+$6015, where both E and W correspond to distinct radio-loud AGN with a projected linear separation of $\sim 70$~pc. This scenario is supported by the flat and inverted spectra of both features (Section~\ref{sec:spix}). However, such objects are very rare \citep[][and references therein]{2018RaSc...53.1211A,2019NewAR..8601525D}, and J2102$+$6015 would be a unique example at high redshift. Moreover, the presence of the weak component C between E and W, and the fact that the characteristic east--west position angle of the structural extensions in the E and W features exactly coincide with the putative SMBHB position angle seem too unlikely. Claiming J2102$+$6015 as a good SMBHB candidate would clearly require further strong independent observational evidence.

\subsection{A young CSO}
\label{sec:cso}
Most observed phenomena can be naturally explained in the framework of the CSO scenario. CSOs usually have stable flux density and GHz-peaked radio spectrum \citep{2012ApJ...760...77A}. In this scenario, the E and W features can be regarded as two terminal hotspots. The location of the nucleus remains uncertain, although the newly identified component C could possibly be associated to the weak radio core. The slow hotspot advance speed is typical for CSOs \citep[e.g.,][]{2005ApJ...622..136G,2012ApJS..198....5A}, consistently with our measurement. CSO hotspots can have complex structures and components with flat or inverted spectra, like in the case of J2102$+$6015. In the CSO scenario, based on the angular separation between E0 and W0, and the proper motion of W0 with respect to E0, we can estimate the kinematic age of the radio source to be 440$^{+400}_{-140}$~years, assuming that the hotspot advance speed remains constant during the early expansion phase of the recently ignited radio AGN.
Indeed, CSOs have typical ages of a few hundred years \citep[e.g. ][]{2012ApJS..198....5A}. 

If the intermediate component C is the core, then the brighter eastern jet is shorter than the fainter western jet. This would be inconsistent with an intrinsic symmetry of the AGN environment and could only be explained by the inhomogeneity of the ISM around the two opposite hotspots. If, in turn, C is not the core but a faint jet knot, and the true core remains undetected with VLBI, the above assumption of an inhomogeneous ISM is not necessary. The brightness temperature of E0 is unusually high for a hotspot, although such a high value is not impossible \citep[e.g. ][]{2006A&A...450..959O,2012ApJS..198....5A}.
It suggests that E0 is an active hotspot and constantly receiving fresh relativistic electrons from the nucleus. This argument is consistent with the flat spectrum of E0 and the young age of the source. 

Another example of a high-redshift ($z=6.12$) CSO candidate is J1427+3312 \citep[][]{2008A&A...484L..39F,2008AJ....136..344M} which has double hotspots separated by 160~pc. J2102$+$6015 has a smaller size and thus younger age than J1427+3312, but is more luminous by two orders of magnitude. The discovery of a population of young radio sources shortly after the end of cosmic reionisation is crucial for understanding the AGN triggering processes and black hole growth in the early Universe. Among the radio-loud high-redshift AGNs, there are an equal number of beamed sources (blazars) and unbeamed sources (CSO or GPS sources) \citep{2016MNRAS.463.3260C}. This over-abundance of CSOs cannot be simply explained by sample selection effects, as the flux density of blazars is amplified by Doppler boosting; at the same flux density limit, blazars are more easily detected. In CSOs, the energy of the jet is mainly deposited in the hotspots. The injected jet power and the hotspot size can constrain the energy density of the particles and  magnetic fields. When the hotspot size is very small (the radio source is very compact), the magnetic field energy density is higher than the cosmic microwave background (CMB) photon energy density, rendering the radiative cooling to be dominated by synchrotron radiation. Such a case may happen in the youngest and smallest CSOs. This mechanism possibly makes the high-redshift CSO hotspots very bright and easily observable. 

VLBI polarimetric observations of AGNs enable us to probe the physical conditions of the interstellar medium within the host galaxies, while polarimetric observations of high-redshift AGNs are very rare so far \citep{2020NatCo..11..143A}. J2102$+$6015 is one of the most luminous radio sources among $z>4.5$ AGNs \citep{2016MNRAS.463.3260C}. Moreover, its flux density  seems stable, unlike in the case of J0906+6930, which recently undergoes a period of fading \citep{2017MNRAS.468...69Z,2020NatCo..11..143A}. So J2102$+$6015 is among the most suitable high-redshift AGN targets for polarimetric VLBI observations. The radio emission of CSOs is known to be depolarized in general, although there are cases with polarized components detected in CSOs \citep[e.g.][]{2005ApJ...622..136G,2016MNRAS.459..820T}. 
Future polarization-sensitive VLBI observations could therefore be valuable to distinguish between the core--jet versus CSO scenarios in J2102$+$6015. Sensitive high-resolution radio imaging at multiple frequencies would help to confirm the existence of component C and its possible association with a CSO core, and to better constrain the spectral properties of the complex mas-scale radio features. Because of the slow apparent proper motion, it is important to continue the 8-GHz imaging of J2102$+$6015 for many years or decades to come. This would be essential for improving the proper motion measurement accuracy.

\section{Summary}
\label{sec:sum}

We reported and analysed new and archival VLBI observations of the high-redshift ($z=4.575$) quasar J2102$+$6015. High-resolution 2- and 8-GHz images from various VLBI projects consistently show a morphology with a compact, bright eastern component and a weaker western component, both having flat or inverted spectra. In our most sensitive image made with the VLBA in 2017, we also found a weak component between the eastern and western features for the first time. It might be the AGN core or a knot in the eastern jet. Comparison of the flux densities obtained from VLBI and low-resolution radio observations indicates almost no diffuse emission extended to arcsec scales. The flux density of the source is fairly stable over a 24-year period. The angular separation speed between the eastern and western features is estimated as $(0.023 \pm 0.011)$~mas\,year$^{-1}$, which yields a kinematic age of about 440~years for this radio source.
The radio spectrum is peaking at around 6~GHz in the source rest frame. The radio properties, including the triple structure, stable flux density, slow proper motion, GHz-peaked spectrum all suggest that the CSO scenario is more favourable compared to others such as a blazar core--jet and SMBHB which lack strong observational evidence. 

The detection of young radio sources at one-tenth of the current age of the Universe is crucial for studying how AGNs are ignited, how feedback impacts the evolution of the host galaxy, and what are the properties of the interstellar environment in the first-generation AGNs. At high redshifts, the compact and highly magnetised hotspots of CSOs are less affected by the enhanced CMB energy density and have the opportunity to be observed in high-resolution radio observations.

\section*{Acknowledgements}

This work used resources of China SKA Regional Centre prototype \citep{2019NatAs...3.1030A}, funded by the National Key R\&D Programme of China (2018YFA0404603) and Chinese Academy of Sciences (114231KYSB20170003). This work was supported by the Hungarian Research, Development and Innovation Office (OTKA K134213 and 2018-2.1.14-T\'ET-CN-2018-00001). The Astrogeo Center Database of brightness distributions, correlated flux densities, and images of compact radio sources produced with VLBI is maintained by L. Petrov. This study made use of an ad-hoc VLBI network which consists of Badary, Seshan25, Svetloe, Yebes, and Zelenchukskaya stations \citep{Frey:2019JI}.
Badary, Svetloe, and Zelenchukskay stations are operated by the Scientific Equipment Sharing Center of the Quasar VLBI Network \citep{SHUYGINA2019150}.

\section*{Data Availability}

The data sets underlying this paper were derived from sources in
the public domain (in case of the Astrogeo database, \url{http://astrogeo.org}) or are available from the corresponding author upon reasonable request.



\bibliographystyle{mnras}
\bibliography{refs} 

\clearpage


\begin{table*}
	\centering
	\caption{Information about the VLBI observations of J2102$+$6015}
	\label{tab:obs}
	\begin{tabular}{cccrcccrccc} 
		\hline
	
		Code & Date & Frequency & Time & Bandwidth & $B_\mathrm{maj}$ & $B_\mathrm{min}$ & $B_\mathrm{P.A.}$ & Peak & $\sigma_{\rm rms}$ & Ref.\\
		 & YYYY-MM-DD & (GHz) & (min) & (MHz)  & (mas) & (mas) & \multicolumn{1}{c}{($\degr$)} & (mJy beam$^{-1}$) & (mJy beam$^{-1}$) &\\
		\hline
		\multicolumn{11}{c}{VLBA session} \\ \hline
		BZ064A & 2017-02-05 & 8.4 & 30 & 512 & 2.5  & 0.8  & $-$1.6  & 97.9 & 0.1 & 1 \\
		\hline
		\multicolumn{11}{c}{Astrogeo archive sessions} \\ \hline
BB023 & 1994-08-12 & 8.3 & 1.5 & 16 & 2.2  & 0.9  & $-$65.4  & 138.4 & 3.1 & 2 \\
... & ... & 2.3 & 3.0 & 16 & 9.6  & 4.4  & $-$38.5  & 238.7 & 2.2 & ... \\
BP133 & 2006-12-18 & 8.6 & 6.5 & 32 & 1.4  & 1.2  & $-$56.7  & 91.4 & 0.7 & 3 \\
... & ... & 2.3 & 6.5 & 32 & 5.0  & 4.3  & $-$46.9  & 181.4 & 0.7 & ... \\
UF001H & 2017-05-01 & 8.7 & 7.2 & 384 & 1.3  & 1.1  & 13.2  & 98.5 & 0.2 & 4 \\
... & ... & 2.3 & 7.6 & 128 & 5.5  & 4.3  & 13.1  & 200.6 & 0.3 & ... \\
UG002J & 2018-06-09 & 8.7 & 5.0 & 384 & 1.5  & 1.1  & $-$14.0  & 94.6 & 0.2 & 4 \\
... & ... & 2.3 & 5.0 & 96 & 5.7  & 4.3  & $-$18.4  & 175.4 & 0.4 & ... \\
\hline
\multicolumn{11}{c}{RUA astrometric sessions} \\ \hline
RUA012 & 2017-02-04 & 8.6 &168 &160 & 1.0 & 0.9 & 70.1 & 102.3  & 0.2 & 5 \\
RUA015 & 2017-04-08 & ... &375 &... & 1.0 & 0.9 & 70.1 & 101.1  & 0.1 & ... \\
RUA016 & 2017-04-22 & ... &315 &... & 0.8 & 0.8 & $-$44.9 & 78.7 & 0.2  & ... \\
RUA017 & 2017-08-19 & ... &336 &... & 1.0 & 0.8 & $-$39.7 & 111.0 & 0.1 & ... \\
RUA018 & 2017-08-26 & ... &384 &... & 1.5 & 0.5 & $-$21.5 & 72.7 & 0.2  & ... \\
RUA023 & 2018-03-31 & ... &304 &... & 0.9 & 0.8 & $-$13.1 & 94.2 & 0.1  & 1 \\
RUA033 & 2019-09-28 & ... &252 &... & 1.1 &0.9  & 48.4    & 107.7  & 0.1 & ...\\
		\hline
		\multicolumn{11}{p{14cm}}{\footnotesize{Notes: Col.~1 -- project code; Col.~2 -- observation date; Col.~3 -- observing frequency; Col.~4 -- on-source integration time; Col.~5 -- bandwidth; Col.~5--7 -- major and minor axes of the synthesized beam (full width at half-maximum, FWHM) and the position angle of the major axis, measured from north to east; Col.~8 -- peak intensity in the image; Col.~9 -- r.m.s noise in the image; Col.~10 -- reference to the corresponding VLBI experiment.}} \\
		\multicolumn{11}{p{14cm}}{\footnotesize{References: 1--This paper; 2--\citet{2002ApJS..141...13B}; 3--\citet{2008AJ....136..580P}; 4--Astrogeo database; 5--\citet{2018A&A...618A..68F}.}}
	\end{tabular}
\end{table*}

\clearpage

\appendix
\section{Figures and tables}
\label{appendix}

\begin{table*}
	\caption{Model-fitting parameters of the VLBI components.}
	\label{tab:parm}
	\begin{tabular}{cccccccc}
		\hline
		Epoch&  Comp. &  $S_\mathrm{peak}$ &  $S_\mathrm{int}$ & $R$ &  P.A. &  $D_\mathrm{comp}$&  $T_\mathrm{B}$\\
		(YYYY-MM-DD)    &    &  (mJy beam$^{-1}$) &  (mJy) &  (mas) &  ($\degr$)& (mas) & ($\times$10$^{10}$K) \\
		\hline
1994-08-12&E0&138.4$\pm$7.7 &145.1$\pm$8.8 &... &... &0.25$\pm$0.03 &22.20$\pm$4.56 \\
2006-12-18&E0&91.2$\pm$4.6 &48.6$\pm$2.4 &... &... &0.31$\pm$0.01 &4.55$\pm$0.25 \\
...&E2&64.6$\pm$3.3 &36.9$\pm$1.9 &0.54$\pm$0.13 &$-$95.7$\pm$0.4 &0.83$\pm$0.01 & ... \\
...&E1&74.7$\pm$3.8 &54.1$\pm$2.8 &0.58$\pm$0.13 &90.2$\pm$0.5 &0.65$\pm$0.01 & ... \\
...&W0&5.4$\pm$0.4 &5.3$\pm$0.5 &9.84$\pm$0.13 &$-$89.2$\pm$0.2 &0.29$^\star$ &... \\ 
2017-02-05 &E0&101.3$\pm$5.1 &121.6$\pm$6.1 &... &... &0.60$\pm$0.01 &3.27$\pm$0.17 \\
...&E1&26.8$\pm$1.4 &9.8$\pm$0.6 &0.79$\pm$0.09 &79.7$\pm$0.3 &0.08$^\star$ & ... \\
...&E2&34.7$\pm$1.8 &25.6$\pm$1.3 &0.79$\pm$0.09 &$-$98.2$\pm$0.3 &0.78$\pm$0.01 & ... \\
...&C&1.3$\pm$0.2 &1.3$\pm$0.3 &4.69$\pm$0.09 &$-$88.2$\pm$0.9 &0.36$^\star$ &$\ge$0.10 \\
...&W0&2.4$\pm$0.3 &5.5$\pm$0.7 &9.94$\pm$0.09 &$-$88.8$\pm$0.3 &1.36$\pm$0.16 &... \\ 
2017-02-04 &E0&90.3$\pm$4.5 &88.9$\pm$4.5 &... &... &0.43$\pm$0.01 &4.45$\pm$0.23 \\
+2017-04-08&E1&53.9$\pm$2.7 &25.3$\pm$1.3 &0.51$\pm$0.09 &97.7$\pm$0.2 &0.63$\pm$0.01 & ... \\
+2017-04-22&E2&45.3$\pm$2.3 &34.2$\pm$1.7 &0.61$\pm$0.09 &$-$101.3$\pm$0.4 &0.78$\pm$0.01 & ... \\
...&W1&1.0$\pm$0.1 &0.8$\pm$0.1 &9.45$\pm$0.08 &$-$89.4$\pm$0.2 &0.27$^\star$ & ... \\
...&W0&1.9$\pm$0.1 &1.6$\pm$0.1 &10.15$\pm$0.08 &$-$89.7$\pm$0.1 &0.20$^\star$ &... \\ 
...&W2&1.1$\pm$0.1 &0.9$\pm$0.1 &10.92$\pm$0.08&$-$90.0$\pm$0.3 &0.34$^\star$ & ... \\
2017-05-01 &E0&102.5$\pm$5.1 &115.3$\pm$5.8 &... &... &0.60$\pm$0.01 &2.87$\pm$0.15 \\
...&E2&51.0$\pm$2.6 &25.5$\pm$1.3 &0.82$\pm$0.11 &$-$102.5$\pm$0.2 &0.77$\pm$0.01 & ... \\
...&E1&31.0$\pm$1.6 &6.2$\pm$0.5 &0.92$\pm$0.11 &80.3$\pm$0.5 &0.13$^\star$ & ... \\
...&W0&3.0$\pm$0.3 &5.2$\pm$0.5 &10.12$\pm$0.11 &$-$89.5$\pm$0.2 &1.06$\pm$0.09& ... \\ 
2017-08-19 &E0&103.5$\pm$5.2 &109.5$\pm$5.5 &... &... &0.43$\pm$0.01 &5.43$\pm$0.28 \\
+2017-08-26&E1&57.0$\pm$2.9 &30.1$\pm$1.5 &0.53$\pm$0.08 &92.1$\pm$0.2 &0.60$\pm$0.01 & ... \\
...&E2&44.5$\pm$2.3 &37.9$\pm$2.0 &0.62$\pm$0.08 &$-$103.7$\pm$0.4 &0.71$\pm$0.01 & ... \\
...&W1&0.9$\pm$0.1 &0.8$\pm$0.2 &9.39$\pm$0.08 &$-$89.0$\pm$0.3 &0.30$^\star$ & ... \\
...&W0&1.1$\pm$0.1 &1.0$\pm$0.1 &10.15$\pm$0.08 &$-$89.9$\pm$0.2 &0.24$^\star$ &... \\ 
...&W2&1.0$\pm$0.1 &1.0$\pm$0.1 &10.95$\pm$0.08 &$-$90.1$\pm$0.2 &0.23$^\star$ & ... \\
2018-03-31 &E0&94.2$\pm$4.7 &98.2$\pm$4.9 &... &... &0.43$\pm$0.01 &4.98$\pm$0.26 \\
...&E1&49.9$\pm$2.5 &26.5$\pm$1.3 &0.52$\pm$0.08 &90.9$\pm$0.1 &0.66$\pm$0.01 & ... \\
...&E2&50.2$\pm$2.5 &28.5$\pm$1.4 &0.54$\pm$0.08 &$-$113.6$\pm$0.1 &0.54$\pm$0.01 & ... \\
...&P1$^*$&10.9$\pm$0.6 &3.6$\pm$0.2 &1.07$\pm$0.08 &$-$82.1$\pm$0.2 &0.08$^\star$ & ... \\
...&P2$^*$&7.1$\pm$0.4 &2.7$\pm$0.2 &1.27$\pm$0.08 &$-$111.3$\pm$0.3 &0.09$^\star$ & ... \\
...&W1&0.9$\pm$0.1 &0.9$\pm$0.1 &9.50$\pm$0.08 &$-$89.5$\pm$0.2 &0.20$^\star$ & ... \\
...&W2&0.6$\pm$0.1 &0.6$\pm$0.1 &10.69$\pm$0.08 &$-$90.2$\pm$0.4 &0.33$^\star$ & ... \\
2018-06-09 &E0&94.2$\pm$4.7 &103.2$\pm$5.2 &... &... &0.58$\pm$0.01 &2.82$\pm$0.15 \\
...&E1&36.3$\pm$1.8 &5.9$\pm$0.3 &0.81$\pm$0.09 &111.1$\pm$0.1 &0.07$^\star$ & ... \\
...&E2&43.4$\pm$2.2 &22.0$\pm$1.1 &0.82$\pm$0.08 &$-$103.2$\pm$0.1 &0.78$\pm$0.01 & ... \\
...&P1$^*$&18.5$\pm$1.0 &5.1$\pm$0.3 &1.09$\pm$0.12 &44.5$\pm$0.4 &0.13$^\star$ & ... \\
...&W2&2.3$\pm$0.2 &2.5$\pm$0.3 &10.70$\pm$0.12 &$-$89.6$\pm$0.3 &0.41$\pm$0.10 &... \\
2019-09-28 &E0&107.6$\pm$5.4 &107.1$\pm$5.4 &... &... &0.41$\pm$0.01 &5.84$\pm$0.30 \\
...&E1&69.9$\pm$3.5 &24.2$\pm$1.2 &0.55$\pm$0.10 &88.2$\pm$0.1 &0.51$\pm$0.01 & ... \\
...&E2&55.7$\pm$2.8 &23.3$\pm$1.2 &0.66$\pm$0.10 &$-$102.7$\pm$0.2 &0.66$\pm$0.01 & ... \\
...&W1&1.1$\pm$0.1 &1.1$\pm$0.1 &9.58$\pm$0.10 &$-$88.7$\pm$0.2 &0.27$^\star$ & ... \\
...&W2&1.0$\pm$0.1 &1.0$\pm$0.2 &10.70$\pm$0.10 &$-$89.6$\pm$0.3 &0.32$^\star$ & ... \\
        \hline
\multicolumn{8}{p{14cm}}{\footnotesize{Notes: Col.~1 -- observing date; Col.~2 -- identifier of the fitted model component; Col.~3 -- peak intensity; Col.~4 -- total flux density; Col.~5 -- angular separation of the component with respect to E0; Col.~6 -- position angle of the component with respect to E0, measured from north to east; Col.~7 -- fitted FWHM size of the circular Gaussian model component; Col.~8 -- brightness temperature. 
}} \\
\multicolumn{8}{p{14cm}}{\footnotesize{$^*$At some epochs, the E feature showed diffuse emission that could not be fully fitted with the circular Gaussians, thus delta-model components P$n$ ($n$=1,2) were used to account for the extended flux density.
}} \\
\multicolumn{8}{p{14cm}}{\footnotesize{$^\star$The size of the component was too small to be fitted with a Gaussian, the minimum resolvable size is used as an upper limit \citep[see][]{2005AJ....130.2473K}.
}}
	\end{tabular}
\end{table*}

\begin{table*}
	\centering
	\caption{Radio observations of J2102$+$6015 from previous surveys and literature.}
	\label{tab:spec}
	\begin{tabular}{ccccc} 
		\hline
		Frequency & Flux density & Instrument & Survey & Reference \\ 
		(GHz)    &  (mJy)  &  &  & \\
		\hline
		0.15 & 158.9$\pm$16.8 & GMRT & TGSS & 1 \\
		0.33 & 273.0$\pm$55.0 & WSRT & WNESS & 2 \\
		1.4 & 343.0$\pm$69.0 & GBT & WB92 & 3 \\
		1.4 & 318.5$\pm$64.0 & VLA & NVSS & 4 \\
		3 & 271.3$\pm$0.4 & VLA & VLASS & 5 \\
		4.85 & 214.0$\pm$19.0 & GBT & GB6 &6 \\
		4.85 & 228.0$\pm$23.0 & GBT & 87GB & 7 \\
		4.85 & 225.0$\pm$45.0 & GBT & BWE & 8 \\
		8.4 & 164.0$\pm$33.0 & VLA & CLASS & 9 \\
		2.3 & 304.7$\pm$45.7 & VLBI-RUA & - & 10 \\
		2.3 & 298.8$\pm$15.2 & VLBA & - & 11,14 \\
		2.3 & 302.3$\pm$15.7 & VLBA & - & 12,14 \\
		2.3 & 348.3$\pm$17.4 & VLBA & - & 13,14 \\
		2.3 & 305.8$\pm$16.2 & VLBA & - & 13,14 \\
		8.3 & 145.1$\pm$8.8 & VLBA & - & 11,14 \\
		8.4 & 151.8$\pm$7.8 & VLBA & - & 14 \\
		8.6 & 158.4$\pm$23.8 & VLBI-RUA & - & 10 \\
		8.6 & 144.8$\pm$7.6 & VLBA & - & 12,14 \\
		8.6 & 163.8$\pm$9.0 & VLBI-RUA & - & 10,14 \\
		8.6 & 178.8$\pm$9.4 & VLBA & - & 13,14 \\
		8.6 & 156.7$\pm$8.1 & VLBI-RUA & - & 14 \\
		8.6 & 161.2$\pm$8.3 & VLBI-RUA & - & 14 \\
		8.6 & 180.3$\pm$9.4 & VLBI-RUA & - & 14 \\
		8.7 & 152.2$\pm$8.1 & VLBA & - & 13,14 \\
		8.7 & 138.7$\pm$7.2 & VLBA & - & 13,14 \\
		\hline
		\multicolumn{5}{p{10cm}}{\footnotesize{References: 1--\citet{2017A&A...598A..78I}; 2--\citet{1997A&AS..124..259R}; 3--\citet{1992ApJS...79..331W}; 4--\citet{1998AJ....115.1693C}; 5--\citet{2020RNAAS...4..175G}; 6--\citet{1996ApJS..103..427G}; 7--\citet{1991ApJS...75.1011G}; 8--\citet{1991ApJS...75....1B}; 9--\citet{2003MNRAS.341....1M}; 10--\citet{2018A&A...618A..68F}; 11--\citet{2002ApJS..141...13B}; 12--\citet{2008AJ....136..580P}; 13--Astrogeo; 14--This paper}}
	\end{tabular}
\end{table*}


\bsp	
\label{lastpage}
\end{document}